\documentstyle[11pt]{article}
\catcode`@=11
\def\@citex[#1]#2{\if@filesw\immediate\write\@auxout{\string\citation{#2}}\fi
  \def\@citea{}\@cite{\@for\@citeb:=#2\do
    {\@citea\def\@citea{,\penalty\@m}\@ifundefined
      {b@\@citeb}{{\bf ?}\@warning
       {Citation `\@citeb' on page \thepage \space undefined}}%
\hbox{\csname b@\@citeb\endcsname}}}{#1}}

\def\citer{\@ifnextchar [{\@tempswatrue\@citexr}{\@tempswafalse\@citexr[]}}
\def\@citexr[#1]#2{\if@filesw\immediate\write\@auxout{\string\citation{#2}}\fi
  \def\@citea{}\@cite{\@for\@citeb:=#2\do
    {\@citea\def\@citea{--\penalty\@m}\@ifundefined
       {b@\@citeb}{{\bf ?}\@warning
       {Citation `\@citeb' on page \thepage \space undefined}}%
\hbox{\csname b@\@citeb\endcsname}}}{#1}}
\catcode`@=12
\def\sVEV#1{\left\langle #1\right\rangle}
\def\ie{\hbox{\it i.e.}{}}

\def\etc{\hbox{\it etc.}{}}
\def\nn{\hspace{2mm}}
\def\sss{\scriptscriptstyle}
\def\MeV{\mbox{\rm MeV}}
\def\GeV{\mbox{\rm GeV}}
\def\eV{\mbox{\rm eV}}
\def\abs#1{\left| #1\right|} 
\def\AGUT{{}\;\;\raisebox{.9ex}{$\times$}\raisebox{-.5ex}%
{$\!\!\!\!\!\!\!\!\sss i=1,2,3$} \,(SMG_i \times U(1)_{\sss B-L,i})}
\def\sleq{\raisebox{-.6ex}{${\textstyle\stackrel{<}{\sim}}$}}

\renewcommand{\thefootnote}{\fnsymbol{footnote}}

\begin{document}
\begin{titlepage}
\setcounter{page}{1}
\hfill
\vbox{
    \halign{#\hfil        \cr
           NBI-HE-01-01    \cr
           hep-ph/0101181 \cr
           } 
      }  
\vspace*{1mm}
\begin{center}
{\Large {\bf A Neutrino Mass Matrix Model with many Quantum charges and No SUSY\footnote[1]{Contribution to The Scandinavian Neutrino Workshop, Uppsala, Sweden, 8-10 February 2001}}\\}
\vspace*{15mm}
\vspace*{1mm}
{\ H. B. Nielsen}\footnote[3]{E-mail: hbech@nbi.dk}
and {\ Y. Takanishi}\footnote[4]{E-mail: yasutaka@nbi.dk}

\vspace*{1cm}
{\it The Niels Bohr Institute,\\
Blegdamsvej 17, DK-2100 Copenhagen {\O}, Denmark}\\

\vspace*{.5cm}
\end{center}

\begin{abstract}
We present a model based on our favourite gauge group
which we call Anti-GUT and which in its extended form to be applied for
neutrinos in the see-saw picture consists in that we have
for each family separately a set of gauge fields as in the
Standard Model plus a gauged $(B-L)$-charge. It may function as
a concrete model manifestation of the type of model suggested by
some general statistical considerations as to what a mass matrix model
is to be to match naturally the rough features of the spectrum
and the baryon asymmetry so nicely obtained in see-saw models
along the Fukugita-Yanagida scheme. 
\end{abstract}
\vskip 4cm

January 2001
\end{titlepage}

\newpage

\renewcommand{\thefootnote}{\arabic{footnote}}
\setcounter{footnote}{0}
\setcounter{page}{2}
\section{Introduction}
\indent

Through several years we have developed a specific model 
\citer{oldantiGUT1,oldantiGUT12} 
for at first the charged particle mass spectrum and mixing angles,
but later in various ways we have attempted to extend it to neutrino mass 
matrices and mixing too \citer{neutrinoAGUT1,NT4}. The 
model we have arrived to through 
various prejudices -- originally starting with some 
statistical mass matrices \citer{colinstatistical1,colinstatistical4} --  
We have made our prejudices especially for the idea of 
explaining the smallness of the three fine structure 
constants \citer{DonPicek1,DonPicek6} by letting the Standard Model
Gauge Group ($SMG$) occur as a diagonal subgroup of one with the
$SMG$ assigned to each family separately,
\begin{equation}
  \label{eq:SMG}
  SMG=SU(3)\times SU(2)\times U(1) = S\,(U(3)\times U(2))\nn.
\end{equation}

It is this as diagonal subgroup embedding of the Standard Model 
group which we call the Anti-GUT model or better extended
Anti-GUT gauge group if the third power of the Standard Model 
is extended with a gauged $(B-L)$-charge for
each family separate, too. For us the history of the diagonal 
subgroup embedding originated from an idea called 
``confusion'' \citer{confusion1,confusion4}, but an $SU(5)^3$ were  
independently considered by Rajpoot \cite{Rajpoot}.

Our model may be characterised by stating that the gauge group
which we postulate under a few restricting rather reasonable
assumptions is the biggest allowed gauge group acting non-trivially
on the known $45$ fermions -- counted as Weyl particles -- of the Standard
Model plus three right-handed neutrinos
(which are to be used as the see-saw neutrinos). The requirement
that it should act non-trivially on these $45+3=48$
Weyl fermions means that our extended Anti-GUT group is
a subgroup, by assumption, of the group of all unitary transformations
$U(48)$ of these 48 Weyl fermions into each other. We could say
that in stead of attempting to make a model for the true gauge
group $G$ beyond the Standard Model we take the attitude that it is already
by far sufficiently ambitious to attempt to guess the true gauge group
with the subgroup $H$ of it not transforming any of the mentioned
48 Weyl fermions but trivially divided out, $G/H$. The least
reasonable of the restrictions imposed  is that we assume there to
be NO unification in the sense that we require that the irreducible
representations of the group $G/H$ of our model, Anti-GUT, should not
unify any irreducible representations of the $SMG$ proper. This 
assumption has a weak justification if
you say that we want to assume only small representations so that
for instance the $45$-dimensional representation of GUT-$SU(5)$
is argued not to be likely to be the truth. The possibility for
unifying irreducible representations of the Standard Model
-- what we wanted to assume not to happen -- namely would imply
$SU(5)$ unification, may be not the full $SU(5)$ GUT but for instance
an $SU(5)$ just for one single family. If it were for the third
family this would predict the at unification scale degeneracy
of the $\tau$-lepton and the bottom quark masses, and that would indeed
be a relatively good prediction, although even for this
prediction some correction from SUSY or other ways might be
needed to fix it to work perfectly. The problem with the Grand Unification
predictions is that they have rather small corrections to fit with
and therefore really have to work well in order for a model
with unification of quark and lepton irreducible representations
like in $SU(5)$-GUT to agree with the experimental masses.
That is why the after our opinion embarrassingly big representation
$45$ of $SU(5)$ constitutes a trouble for the GUT and so there can be
a point in avoiding unification, which was our most unreasonable
assumption.

What we consider a more reasonable assumption is that there shall be anomaly
cancellations both for mixed anomalies -- the ones involving also coupling
to gravitons -- and for gauge anomalies without using Green-Schwartz
mechanism \cite{GS}. One can at least say that in the cancellation of 
gauge anomalies
in the pure Standard Model there is no need for other terms than the
ones from the Weyl fermions, so one can at least say that the Standard
Model itself does the job without any Green-Schwartz mechanism. 

But the real assumption important for our model is that the
group should be maximal inside the $U(48)$. It is thus quite
important if one could make some estimate of how many gauge group
generators there should be in the theory behind the Standard Model.
It is a slightly new part of the present talk to point out two
possibilities for making such an estimate:%
\begin{itemize}
\item We can look for how much the general scale of quark and lepton
masses is suppressed relatively to the weak scale compared to the
spread in these masses.
\item We could extract a similar parameter from the baryon asymmetry
fitting in the see-saw neutrino scheme. 
\end{itemize}

In the next section and section three we shall associate this ratio of 
the width
of the peak to the average distance and thereby the effective number
of -- taken as abelian -- mass protecting charge species partly by
making statistics on the mass spectra partly by use of baryon asymmetry.
Then in the fourth section we present our specific model, especially
being specific w.r.t. the quantum number assignments to the various
Higgs fields used to break down our favourite Anti-GUT gauge group
to the Standard Model group. In the fifth section we point out
how our model manage to obtain a good fit to neutrino oscillation
fitted masses and mixing angles for the neutrinos. In the sixth section
we describe the relation of our model to SUSY, we have no
SUSY, but it is only a couple of points that makes this point in our
model needed. In section seven we present how the relation of the
quantum numbers and the anomaly cancellations makes number of 
independent quantum number types appreciably less than the full
number of quantum number types. Finally in section eight we 
conclude that we have a model
that fits both baryogenesis and the mass and mixing spectra very well
and that it is suggested to have a number of effective mass protecting
charges not totally different from the rather high number in our model.

\section{Estimating the effective number of mass suppressing charges}
\indent

In our model is the assumption that the order of magnitude of the
various matrix elements in the mass matrices in the proto-flavour
basis are given by suppression factors from the Higgs fields needed to
be used in order to provided the quantum number transition between
the quantum numbers of the right- and the left-handed Weyl components
connected with the matrix element in question. For, for instance,
the quarks or the charged leptons one first of all need to
have the transition of the quantum numbers associated with the
Standard Model -- the weak hypercharge and the weak isospin. This
is what is achieved by use of the Weinberg-Salam Higgs field. In
our model we have to assign this effect to a Weinberg-Salam Higgs
field having not only the usual quantum numbers but also some
quantum number assignment under the full gauge group proposed
in our case the Anti-GUT model. But in general the quantum numbers
of the model-Weinberg-Salam-Higgs field is not sufficient to provide
most of the quantum number transitions needed to provide the
matrix elements to be non-zero. We therefore need a series of
Higgs fields, which we have given names: $W$, $T$, $\xi$, $\chi$ and $S$.
These fields having nonzero expectation values have to be assigned though
such quantum numbers of course that they do not break the subgroup
of the Anti-GUT group supposed to survive being identified with
the Standard Model Group.

In general a matrix element will need several of these Higgs fields
in order to be nonzero, the more they need the smaller a given
matrix element will be and the smaller the mass getting its main contribution
from that element.

Even if one does not want to trust a special model, even if well
fitting, one could try to ask the statistical question:
Is it likely if we want to fit with a model of this type that we
shall have very few quantum numbers effectively working to make
the suppressions of the matrix elements or is it better to have rather
many such quantum numbers?

We shall here seek to give phenomenologically an estimate of the
effective number of ``statistically independent'' such quantum numbers
(thinking of them for simplicity as abelian quantum numbers).

Very crudely  we might assume that the matrix elements get suppressed
by a factor that is given as the exponential of a distance in the
space of quantum numbers -- provided with an appropriate metric
for this purpose -- between the quantum numbers of the right-handed
to those of the left-handed counted in a ``space of quantum numbers''.

Depending on the number of dimensions $n$ in this ``space of quantum numbers''
the width of the peak of the distance distribution relative to the
average distance is estimated -- using for being able to get a quite
definite number the assumption that the  distribution of the quantum number
combinations is an $n$-dimensional Gaussian distribution around the origo --
to be:
\begin{eqnarray}
\frac{\sqrt{\left\langle(dist - \sVEV{dist})^2\right\rangle}}{\sVEV{dist}}&=&\sqrt{\frac{\Gamma(\frac{n+2}{2})\,
\Gamma(\frac{n}{2})}{\Gamma(\frac{n+1}{2})^2} -1}\label{eq1}\\
&\approx& \sqrt{\frac{1}{n+1}} \qquad {\rm for~large}\label{eq2}\nn n \nn.
\end{eqnarray}

\section{Parameters of the Baryon Asymmetry Calculation}
\indent

It is one of the remarkable somewhat mysterious features of the system
of parameter values in the Standard Model -- and thus a reason for going
beyond it -- that Yukawa couplings are in general rather small numbers
of actually quite varying orders of magnitudes. Indeed the Yukawa
couplings found were so
much less than unity that when at the end
it was found that the top-quark Yukawa coupling was of order unity
it came as a surprise, most of us having expected a lighter top quark.
This means that the majority of quark and charged lepton masses are
indeed, on a logarithmic scale say, distributed around a central
value corresponding to smaller masses than the top-mass which
has a weak scale mass (because of its Yukawa coupling being
of order unity, actually surprising close to just unity \cite{Barshay}).

The simplest way to make a quantitative estimate of how much
the general suppression is to use the average charged lepton and
quark mass logarithms compared to the mass scale given by the weak
interactions, which can be taken to be the mass scale for a quark
or charged lepton in the Standard Model having unit Yukawa coupling, 
\ie\ compared to the top-quark mass. In the same scheme we may study
the average spread of the charge lepton and quark mass logarithms,
the latter being mostly dominated by the spread between the three
families. The parameter that should now tell us about the effective
number of mass protecting -- counted as abelian -- charges is the ratio
of this spread to the suppression average of the logarithms of
masses in the weak interaction unit. In the philosophy that
the Yukawa couplings are really running couplings with say the
Planck scale or some unifying scale close to the latter as the most
``fundamental'' one, we use for our estimated a factor $3.5$ increase
of the charged lepton masses together with the quark masses referred to
a scale at which to define the Yukawa couplings of $1~\GeV$. This is
namely supposed to correspond to comparing the Yukawa couplings
at the Planck scale.

The logarithmically averaged masses for the three families of 
particles corrected this way are 
\begin{eqnarray}
\exp\sVEV{\ln\,m}_1 &=& 4.0 ~\MeV \qquad( 1.14 ~\MeV \; \hbox{as lepton})\\
\exp\sVEV{\ln\,m}_2 &=& 469~\MeV \qquad( 133 ~\MeV \; \hbox{as lepton})\\ 
\exp\sVEV{\ln\,m}_3 &=& 19 ~\GeV \qquad( 5.5 ~\GeV \; \hbox{as lepton}) \nn,\\
\end{eqnarray}
where the lower indices $(i=1,2,3)$ denote family numbers.  The 
logarithmic average for the whole set of all the nine 
charged quarks and leptons is 
\begin{equation}
\exp\sVEV{\ln\,m}_{\rm total} = 320 ~\MeV \qquad( 94  ~\MeV \; \hbox{as lepton})\\\nn.
\end{equation}
The root mean square spread of the logarithms of the 
family average masses is estimated as - remembering to divide only
by $2$ rather than $3$ - 
\begin{equation}
\frac{1}{2}\,\sum_{i=1}^3 (\sVEV{\ln\,m}_i - \sVEV{\ln\,m}_{\rm total})^2 = 18
\end{equation}
corresponding to a one standard deviation spread in the family averages 
of $\exp \sqrt{18} =$ a factor $70$. To get the full mean square
spread we should add to the $18$ the mean square ``fluctuation'' inside
the families. By far the biggest intra family mean square fluctuation 
comes from the third family and is $7._5$, the rest is about $2._5$. So the 
root mean 
square spread in the logarithm for the charged quark and leptons is 
$\sqrt{28} =5._2$, corresponding to a factor $\exp(5._2) = 200$.

On the other hand the logarithmic average $320~\MeV$ is about a factor $560$
under the weak scale mass corresponding to $6._3$ for the logarithm.
The ratio of this $6._3$ to the $5._2$ equal to $1.2$ is a measure for how
much the masses are suppressed generally compared to their spread. 
From this one it should be possible -- with some corrections though -- to
extract an estimate of the effective number of ``independently assigned'' 
approximately conserved charges needed to suppress the masses.

One correction consists in that the matrix elements most likely to give
the masses - which were the ones on which we just made statistics - are
relatively big because they otherwise tend to be dominated out, so that we 
must estimate that the average over the matrix elements in the mass matrices
must be significantly more suppressed than the masses. With the masses making 
up half the number of matrix elements one may estimate this increase in
the numerical suppression logarithm for the matrix elements relative to the 
masses to be $1/\sqrt{\pi}= 0.56$ times the width of the distribution,
that itself will be the same for the masses and the matrix elements in general.
If the mass matrices are not symmetric as for neutrinos there are rather 
three times as many general matrix elements as masses and we may estimate 
to use $0.8$ rather than the $.56$.

We end up estimating that the inverse of the quantity (\ref{eq1},\ref{eq2}) 
shall be 
\begin{equation}
\frac{\sVEV{dist}}{\sqrt{\left\langle(dist - \sVEV{dist})^2\right\rangle}}=
\frac{6._3}{5._2} +0.8 = 2.0\qquad (\hbox{or} \nn \frac{6._3+\ln{3.5}}{5._2} +0.8 = 2.2_5 \nn\hbox{as leptons}).
\label{estimate}
\end{equation}
This means that if we use (\ref{eq2}) we estimate that the number 
of statistically independently used (as if abelian) charge types 
in a model aiming at the experimental spectrum should be 
\begin{equation}
n = 2.0^2 -1 =3 \qquad (\hbox{or} \nn {2.2_5}^2 + 1 = 6\;\; \hbox{as leptons})
\end{equation}
but using the more accurate (\ref{eq1}) we rather get 
\begin{equation}
n=2.2  \qquad (\hbox{or} \nn 2.7 \; \hbox{as leptons})
\end{equation}
so that one can be a bit surprised almost that our model 
\citer{oldantiGUT12,neutrinoAGUT2} in which only four abelian 
$U(1)$'s play the role of importance 
for charged spectra should fit well. The ideal model should only have 
about $2.2$ types of charges important. It should in this
connection then be remarked that because we have one Higgs field, namely $S$,
with expectation value of order unity, we have really missed one charge 
type effectively for that reason. So we have $4-1=3$ effective 
charges; that agrees
with the estimate. But a lot of our charges are really not ``effective'' in 
the suppression: The non-abelian ones follow the abelian ones and 
are not independent, and the $U(1)_{B-L,i}$ charges were not needed.

\subsection{Baryogenesis $CP$ violating parameter}

A second way to estimate very much the same parameter is the 
$CP$-violating parameter $\epsilon$ that gives the relative excess 
production of lepton number  
for the $CP$ violation in the decay of the right-handed neutrino 
functioning as a see-saw neutrino 
\begin{equation}
  \label{eq:epsilonCP}
 \epsilon^{N_R}_{\ell}\equiv{\Gamma_{N_R\ell}-\Gamma_{N_R\bar\ell}
\over \Gamma_{N_R\ell}+ \Gamma_{N_R\bar\ell}}\nn, 
\end{equation}
where $\Gamma_{N_R\ell}\equiv \sum_{\alpha,\beta}\Gamma(N_R\to \ell^\alpha
\phi_{\sss WS}^\beta)$ and 
$\Gamma_{N_R\bar\ell}\equiv \sum_{\alpha,\beta}\Gamma(N_R\to
\bar\ell^\alpha \phi_{\sss WS}^{\beta \dagger})$ are the $N_R$
decay rates (in the $N_R$ rest frame), summed  over the
neutral and charged leptons (and Weinberg-Salam Higgs fields) 
which appear as final states in the $N_R$ decays.
   
Since such $CP$-violation parameters are in the scheme of lepton violation 
caused baryon asymmetry in the early universe (Fukugita-Yanagida scheme 
\cite{FY}) the important factor in obtaining the baryon number to entropy 
ratio they are phenomenologically accessible. The reason that we claim 
it is so similar -- in a statistical way -- to the just mentioned spread 
to overall suppression ratio (in the log of the masses) is seen from 
the expression in terms of Yukawa couplings or essentially equivalently 
the Dirac mass matrix for neutrinos, which is proportional to the couplings
of the right-handed (= see-saw) neutrinos \cite{see-saw1,see-saw2} 
to the left-handed (= in practice observed) neutrinos \citer{Luty,CRV}
 \begin{equation}
\epsilon_i = \frac{2}{8 \pi \sVEV{\phi_{\sss WS}}^2 ((M_\nu^D)^{\dagger} M_\nu^D)_{ii}}\sum_{j\not= i} {\rm Im}[((M_\nu^D)^{\dagger} M_\nu^D)^2_{ji}] \left[\, f \left( \frac{M_j^2}{M_i^2} \right) + g \left( \frac{M_j^2}{M_i^2} \right)\,\right] 
\end{equation}
where $f$ comes from the one-loop vertex contribution and
$g$ comes from the self-energy contribution, which
can be calculated in perturbation theory 
only for differences between Majorana neutrino masses which 
are sufficiently large compared to its decay widths, \ie\ 
the mass splittings satisfy the condition, 
$\abs{M_i-M_j}\gg\abs{\Gamma_i-\Gamma_j}$:
\begin{eqnarray}
f(x) &=& \sqrt{x} \left[1-(1+x) \ln \frac{1+x}{x}\right]\nn, \label{eq:vas1}\\
g(x) &=& \frac{\sqrt{x}}{1-x} \label{eq:vas2}\nn.
\end{eqnarray}

It is namely easily seen that the lower the overall size of the 
Yukawa couplings the smaller will be the $CP$-violating parameter $\epsilon$.
So by seeking to extract $\epsilon$ phenomenologically we obtain a knowledge 
about the Dirac mass matrix for the neutrinos much similar to what we
get from the mass suppression compared to the weak scale for the
charged particle mass matrices.

One might attempt to make a statistical estimation of the 
relation of the spread ratio and the $\epsilon$-order of 
magnitude by statistical assumptions, but it is presumably 
better to use an in the details concrete model that fits 
and is of the type we imagine, \ie\ a model in which it is 
assumed that all couplings are fundamentally of order unity 
and only because of approximately conserved quantum numbers 
that need some Higgs field vacuum expectation values (which 
may be very small) for their breaking there can appear small 
numbers in the model at all.   

\subsection{Baryogenesis via lepton number violation}
\indent

The physics of the baryon excess creation in big bang via 
the lepton number violation runs in the following way \cite{KT}: %

In the models like ours having see-saw neutrinos we get 
at first an excess of $B-L$ which is conserved -- in the 
Standard Model even at temperatures so high that sphalerons 
\citer{'tHooft1,sphaleron} allows
violation of baryon number $B$ and Lepton number $L$ -- as an 
``accidental'' symmetry.

As the temperature during big bang falls and passes the order of 
magnitudes of the three see-saw neutrino masses they one after the other
gets out-of-equilibrium by being more copious than they should be in 
equilibrium. The excess decays in a way that violates $CP$ a little
amount given by the parameter $\epsilon$ and thereby leaves an over
abundance of -- actually at first a lepton number under abundance --
of $(B-L)$-charge. As finally temperatures get so low that the 
see-saw neutrinos get out of significance -- even virtually --
the amount of $(B-L)$ gets logged in and cannot change anymore.

There is a significant dilution or wash out of the $(B-L)$-quantum 
number arising from the two lightest of the three see-saw neutrinos 
it turns out in our model because their life times are short 
compared to the Hubble times at their time of going 
out-of-equilibrium.

One could now fear that the breaking of $(B-L)$-conservation caused by
two light see-saw neutrinos could also make dilution or disappear the 
$(B-L)$ amount produced by the heaviest see-saw neutrino. Luckily for 
the success of our model, however, it can be argued that the 
super-position of flavours of the leptons (at first left-handed)
produced by the decays of the heaviest see-saw neutrinos are 
quantum mechanically orthogonal to those super-positions which 
can be diluted by means of the lighter see-saws. Thus it is 
really as if there were three different $(B-L)$-quantum 
numbers separately conserved. Only after $2$-by-$2$ scatterings 
using $t\,t\,\phi_{\sss WS}$ coupling become important there is any 
charge of converting the one flavour $(B-L)$ into the other one.

Thus the separate see-saw neutrinos can only dilute their own 
decay products. Since in our model the heaviest see-saw 
neutrino turns out to have life time accidental very close
order of magnitude-wise to the Hubble time scale at its going 
out-of-equilibrium time, the wash out or dilution of the decay 
products from this heaviest see-saw neutrino is not so significant. 
Thus to very crudest approximation the $(B-L)$-asymmetry is simply 
obtained as the number of heaviest flavour see-saw neutrinos 
multiplied by $\epsilon$. Because of the equilibrium under 
conversion of $-1$ lepton $+1$ baryon, it will be 
$B \sim (B-L)/3$ \cite{HT,API}.

\section{Our Model and fitting}
\indent

Our Anti-GUT model -- extended as well as original -- is
of the type mentioned in the end of subsection $3.1$
as far as a basic assumption in it
is that at the fundamental scale taken to be the Planck scale 
all couplings are of order unity and really are in the calculations
which we perform treated as random numbers of order unity. Then 
we let the results be calculated time after time using these random
numbers -- \ie\ random coupling constants -- fluctuating around zero
in the complex plane with numerical values fluctuating around unity
and at the end we average over the many random number combinations 
and typically we take a logarithmic average of the quantity which we want
to compute. 

The only small quantities are the vacuum expectation values of a 
series of Higgs fields -- $W$, $T$, $\xi$, $S$, $\chi$ with 
expectations values in vacuum 
of the order of one to two orders of magnitude below the 
fundamental scale, and the Higgs fields $\phi_{B-L}$ giving 
the see-saw neutrino scale, and a Higgs field taking over the role 
of the Weinberg-Salam Higgs field $\phi_{\sss WS}$ -- the abelian part of 
the quantum numbers of which are 
presented in the table. (see how to get the representations for the 
non-abelian invariant subgroups following rule of the section seven 
from the table, too.)


\begin{table}[!t]
\caption{All $U(1)$ quantum charges in extended Anti-GUT model.}
\vspace{3mm}
\label{table1}
\begin{center}
\begin{tabular}{ccccccc} \hline\hline
& $SMG_1$& $SMG_2$ & $SMG_3$ & $U_{\sss B-L,1}$ & $U_{\sss B-L,2}$ & $U_{\sss B-L,3}$ \\ \hline
$u_L,d_L$ &  $\frac{1}{6}$ & $0$ & $0$ & $\frac{1}{3}$ & $0$ & $0$ \\
$u_R$ &  $\frac{2}{3}$ & $0$ & $0$ & $\frac{1}{3}$ & $0$ & $0$ \\
$d_R$ & $-\frac{1}{3}$ & $0$ & $0$ & $\frac{1}{3}$ & $0$ & $0$ \\
$e_L, \nu_{e_{\sss L}}$ & $-\frac{1}{2}$ & $0$ & $0$ & $-1$ & $0$ & $0$ \\
$e_R$ & $-1$ & $0$ & $0$ & $-1$ & $0$ & $0$ \\
$\nu_{e_{\sss R}}$ &  $0$ & $0$ & $0$ & $-1$ & $0$ & $0$ \\ \hline
$c_L,s_L$ & $0$ & $\frac{1}{6}$ & $0$ & $0$ & $\frac{1}{3}$ & $0$ \\
$c_R$ &  $0$ & $\frac{2}{3}$ & $0$ & $0$ & $\frac{1}{3}$ & $0$ \\
$s_R$ & $0$ & $-\frac{1}{3}$ & $0$ & $0$ & $\frac{1}{3}$ & $0$\\
$\mu_L, \nu_{\mu_{\sss L}}$ & $0$ & $-\frac{1}{2}$ & $0$ & $0$ & $-1$ & $0$\\
$\mu_R$ & $0$ & $-1$ & $0$ & $0$  & $-1$ & $0$ \\
$\nu_{\mu_{\sss R}}$ &  $0$ & $0$ & $0$ & $0$ & $-1$ & $0$ \\ \hline
$t_L,b_L$ & $0$ & $0$ & $\frac{1}{6}$ & $0$ & $0$ & $\frac{1}{3}$ \\
$t_R$ &  $0$ & $0$ & $\frac{2}{3}$ & $0$ & $0$ & $\frac{1}{3}$ \\
$b_R$ & $0$ & $0$ & $-\frac{1}{3}$ & $0$ & $0$ & $\frac{1}{3}$\\
$\tau_L, \nu_{\tau_{\sss L}}$ & $0$ & $0$ & $-\frac{1}{2}$ & $0$ & $0$ & $-1$\\
$\tau_R$ & $0$ & $0$ & $-1$ & $0$ & $0$ & $-1$\\
$\nu_{\tau_{\sss R}}$ &  $0$ & $0$ & $0$ & $0$ & $0$ & $-1$ \\ \hline \hline
$\phi_{\sss WS}$ & $\frac{1}{6}$ & $\frac{1}{2}$ & $-\frac{1}{6}$ & $-\frac{2}{3}$ & $1$ & $-\frac{1}{3}$ \\
$S$ & $\frac{1}{6}$ & $-\frac{1}{6}$ & $0$ & $-\frac{2}{3}$ & $\frac{2}{3}$ & $0$ \\
$W$ & $-\frac{1}{6}$ & $-\frac{1}{3}$ & $\frac{1}{2}$ & $\frac{2}{3}$ & $-1$ & $\frac{1}{3}$ \\
$\xi$ & $\frac{1}{3}$ & $-\frac{1}{3}$ & $0$ & $-\frac{1}{3}$ & $\frac{1}{3}$ & $0$ \\
$T$ & $0$ & $-\frac{1}{6}$ & $\frac{1}{6}$ & $0$ & $0$ & $0$ \\
$\chi$ & $0$ & $0$ & $0$ & $0$ & $-1$ & $1$ \\
$\phi_{\sss B-L}$ & $-\frac{1}{6}$ & $\frac{1}{6}$ & $0$ & $\frac{2}{3}$ & 
$-\frac{2}{3}$ & $2$\\ \hline
\hline
\end{tabular}
\end{center}
\end{table}
\noindent

That is to say that all the suppressions of the Yukawa couplings 
from the {\it a priori} unity order of magnitude need some mass protecting 
quantum numbers that are different on the right- and left-handed Weyl components
in such a way that it is not just taken up by the Weinberg-Salam Higgs field
but needs to interact with yet another Higgs field to get rid of the charge 
in question, or typically several because that is how we get often very big 
suppressions. 

The model which we present via the table and the statement 
that we have the $39=3 \cdot 13=3 \cdot (12+1)$ gauge fields 
corresponding to the gauge group 
\begin{equation}
  \label{eq:AGUTgauge}
  \AGUT \nn,
\end{equation}
where $SMG_i$ denotes $SU(3)_i\times SU(2)_i\times U(1)_i$ (SM gauge group),
and $i$ denotes the generation, is of course rather arbitrary, at least
in as far as the quantum numbers for the Higgs fields in the table 
are gotten by seeking to the fit the mass spectra just.
 
\subsection{Arguing for our model}

You could, however seek to argue this way: %
Since we get as we shall see a very successful fit to really 
all the parameters in the Standard Model Fermion spectra and
even to the Baryon asymmetry we could say that a model -- namely our model --
with the for our model so characteristic high number of gauge fields 
does fit well, especially it fits well the in foregoing section 
mentioned quantities connected with the general suppression or smallness of 
the Yukawa couplings to the Weinberg-Salam Higgs field compared say to the 
spread the strength of these couplings. But that then means the information 
that there shall be a high number of effective mass protecting approximately
conserved quantum numbers in a model that naturally should fit the statistical
features of the mass spectra and the baryon number in the Universe. 
But then we can switch on to the characterisation of our
gauge group as the one with the maximal number of dimensions still obeying 
the in the introduction mentioned assumptions.  

In addition to the gauge and Higgs and Fermion fields which we have 
explicitly mentioned via the gauge group and the table it is meant in our 
model that once you ask for particles which have masses of the order of the 
``fundamental scale'' (which we think of as the Planck scale) there are 
practically all the not mass protected particles you may ask for and they 
all couple with coupling constants of order unity. This is of course a 
slightly vague and a strong assumption, too, but it is much more likely 
to be true than any specific assumption about the physical fields or 
particles at the Planck scale. The in simplicity competing assumptions
could be that there no Planck mass particles at all, or some string theory say.
To the first proposal -- no particles at all -- we might answer that we
need at least some intermediate fermions to establish effective Weinberg-Salam
Higgs Yukawa couplings which in reality depends on and are proportional 
to some vacuum expectation value that can explain its smallness. So at least 
some such fermions are called for. To the second -- the string theory -- we
might answer that in principle the superstring theories are practically 
looked upon not so far from our assumption ``everything can be found at the 
fundamental scale, and with coupling unity''. Indeed there are infinitely many 
string states and even the lowest levels are for practical purposes reasonably
complicated and the attitude of treating them a bit like random couplings
may not be so unattractive. Usually it is taken that there is a string 
coupling that is small, but that would for instance be based on the 
phenomenological equating of it with the gauge couplings experimentally 
accessible or it would some theoretical justification which might
be somewhat analogous to our ``only the vacuum expectation values are small''
attitude if one takes the note that strictly
speaking in superstring theories \cite{Witten1,Witten2} the string coupling 
is connected with the dilation expectation value in vacuum and thus
is not really a fundamental coupling that is just input into the theory.
Taking, however, for granted that the Standard Model gauge groups
lie inside the full Anti-GUT model as diagonal subgroups -- as would be 
suggested by the many gauge charge argument via our Gauge group -- 
 it is unavoidable that the gauge couplings for the family specific
subgroups be of the order of a square root three stronger than the
corresponding Standard Model and then they are not so small anymore
if one think of the $g$ notation rather than $\alpha$-notation. 
In fact \citer{DonPicek1,DonPicek3} we have long argued that these 
couplings could be 
fitted with phase transition couplings, and that would suggest that at the
end they can be considered of order unity. That would mean that accepting the
gauge group and the embedding of the Standard Model one in the way 
of our model, you would loose the phenomenological suggestion for the 
weak couplings at the fundamental scale, rather the ``everything of order 
unity'' hypothesis gets at least somewhat closer.

Although our model in principle has infinitely many parameters in the sense
that there are practically infinitely many order one fields and couplings
we get by our order of unity assumption and the rule of averaging with
taking them as random numbers rid of these parameters and there remains 
only the vacuum expectation values of our seven Higgs fields from the table.   
 This makes of course our model much more predictive, but the price is
that we then always have these order unity factors which we do not 
really know and all our results can only be taken as order of magnitude 
predictions! 

\subsection{Certain technical points, predicting even the uncertainty}
\indent

Let us mention that really the only order of magnitude predictions are 
not so enormously uncertain again or rather it is even possible to
essentially predict the precise amount  of this uncertainty! The point is 
that with the philosophy that at the Planck scale everything 
exists and has coupling unity there will for a large number of charges
needed to be broken in general be many ways, many Feynman diagrams via
which the transition can be achieved. To make as good an estimate of the
mass matrix elements as possible inside our model assumptions \cite{douglas} 
we have attempted to count these Feynman diagrams and
used a philosophy of them contributing with random phases to make a random
walk estimate of a correction for the number of ways a transition can go,
and these corrections have been included in the fits. That there are 
in many cases rather many Feynman diagrams contributing to the 
same transition, \ie\ the same mass matrix element say, means that 
as they have the same order of magnitude (because they violate the same 
quantum numbers) they add up to a number which has a Gaussian distribution
according to the central limit theorem. This distribution is centered 
around zero -- in the complex plane --  and has a width given by the 
suppression factors and a factor given by the square root of the 
number of Feynman diagrams. These square root of the estimated number
of Feynman diagram correction factors tend to be expressed in terms of 
factorials of the number of vacuum expectation values used and we therefore 
have sometimes called this correction ``factorial correction''. 
But now on general grounds the distribution became all the time
practically a Gaussian one so that the logarithm is distributed 
just in the way a logarithm of such a Gaussian distribution is distributed
and that means with a specific width of $0.64=64\%$. This is why we 
claim that even the uncertainty of the prediction from
our type of model is predicted. If a quantity depends in a multiplicative way
on several matrix elements of course these $64\%$ fluctuation is
increased by a factor that is the square root of the number of matrix elements
multiplied (or divided) together. 

\subsection{Results}

The development of our model w.r.t. quark and lepton masses went 
from first fitting the charged quark and lepton masses and mixing
angles using as parameters the expectation 
values of the Higgs fields $W$, $T$, $\xi$, $S$ and $\phi_{WS}$.
There is some possibilities for playing with the quantum numbers in the
model by adding to the other fields the quantum number combination 
(found in the table) of the field $S$, because this field gets an
expectation value in the fitting that is very close to the fundamental
scale so that there is really a strong violation of the conservation 
of this combination of quantum numbers. Because of this feature of 
our fit, the Higgs field $S$ giving essentially no suppression, getting
in an extra $S$ field into a matrix element is only a small
correction partly via the $S$ factor partly via changing the ``factorial''
corrections mentioned in foregoing subsection. A typical fit 
to the charged masses and mixing angles is presented in table $2$.

\begin{table}[!t]
\caption{Typical fit including averaging over ${\cal O}(1)$ factors. 
All quark masses are running masses at $1~\GeV$ except the top quark 
mass which is the pole mass.}
\begin{displaymath}
\begin{array}{c|c|c}
& {\rm Fitted} & {\rm Experimental} \\ \hline \hline
m_u & 3.1 ~\MeV & 4 ~\MeV \\
m_d & 6.6 ~\MeV & 9 ~\MeV \\
m_e & 0.76 ~\MeV & 0.5 ~\MeV \\
m_c & 1.29 ~\GeV & 1.4 ~\GeV \\
m_s & 390 ~\MeV & 200 ~\MeV \\
m_{\mu} & 85 ~\MeV & 105 ~\MeV \\
M_t & 179 ~\GeV & 180 ~\GeV \\
m_b & 7.8 ~\GeV & 6.3 ~\GeV \\
m_{\tau} & 1.29 ~\GeV & 1.78 ~\GeV \\
V_{us} & 0.21 & 0.22 \\
V_{cb} & 0.023 & 0.041 \\
V_{ub} & 0.0050 & 0.0035 \\
J_{\sss CP} & 1.04\times10^{-5} & 2\!-\!3.5\times10^{-5}\\ \hline \hline
\end{array}
\end{displaymath}
\label{MassenmitFF}
\end{table}

Since we use the Higgs field $\phi_{B-L}$ to fit the overall scale of the
see-saw neutrino masses and thereby the overall scale for the neutrino
oscillations \citer{totsuka,valle} we can genuinely only hope to predict 
the mass ratio of the two mass magnitudes involved in respectively 
the atmospheric neutrino
oscillations $\sqrt{\Delta m^2_{\rm atm}} = \sqrt{ 3.2 \times 10^{-3}~\eV^2} 
= 5 \times 10^{-2}~\eV$ and the solar one $\sqrt{\Delta m^2_{\odot}} 
=\sqrt{5 \times 10^{-6}~\eV^2} = 2 \times 10^{-3}~\eV$, \ie\ we can only 
hope to predict the ratio $\Delta m^2_{\rm atm}/\Delta m^2_{\odot}
=1.5 \times 10^{-3}$. Indeed our best fit turns out to give for this
ratio the prediction $5.8 \times 10^{-3}$, and that shall be considered a 
great success since our expected uncertainty is 
$\pm 64\%\cdot\sqrt{2^2+2^2}=181\%$ corresponding to uncertainty
w.r.t. $\exp(1.81)=5$ to $1/\exp(1.81)=0.2$. With these theoretical 
uncertainty we get 
\begin{eqnarray}
  \label{eq:bestresults}
\frac{\Delta m_{\odot}^2}{\Delta m_{\rm atm}^2} \!\!&=&\!\! 5.8{+30\atop-5}\times10^{-3}\\
  \tan^2\theta_{\odot}\!\!&=&\!\! 8.3{+21\atop-6}\,\times10^{-4}\\
  \tan^2\theta_{e3}\!\!&=&\!\! 4.3{+11\atop-3}\,\times10^{-4}\\
  \tan^2\theta_{\rm atm}\!\!&=&\!\! 0.97 {+2.5\atop-0.7}\\
  Y_B \!\!&=&\!\! 1.5 {+20\atop-1.4}\times 10^{-11}\nn.
\end{eqnarray}
corresponding the experimental data:
\begin{eqnarray}
  \label{eq:expresul}
{\frac{\Delta m_{\odot}^2}{\Delta m_{\rm atm}^2}}_{\sss, \rm exp} \!\!&=&\!\! 1.5 {+1.5\atop-0.7}\times10^{-3}\\
  {\tan^2\theta_{\odot}}_{\sss, \rm exp}\!\!&=&\!\! (0.33-2.5)\times 10^{-3}\\
  {\tan^2\theta_{e3}}_{\sss, \rm exp}\!\!&\sleq&\!\! 2.6\times10^{-2} \\
  {\tan^2\theta_{\rm atm}}_{\sss, \rm exp}\!\!&=&\!\! (0.43 - 4.2)\\  
  {Y_B}_{\sss, \rm exp} \!\!&=&\!\! (2-9)\times 10^{-11}\nn.
\end{eqnarray}

\section{The compensation removing much of neutrino hierarchy}
\indent

Now we have a model one may ask how we solve in that model some of 
the strange features of the neutrino oscillation parameters:

For instance how does it come that the mass ratio of the two 
neutrino oscillation masses is relatively small compared to 
similar ratios in the charged mass matrices, namely a factor $10$ to $30$
in mass ratio compared to rather a factor $100$ for the family spacing in
the charged cases. And also this is really mysterious from the 
point of view of the models we have in mind: How can the two heaviest 
left-handed neutrinos have after all a hierarchical mass ratio
even if only factor $10$ to $30$ when the atmospheric mixing angle which
would suggestively be related to the two heaviest of the left-handed neutrinos
is of order unity? 

The answer in our model goes like this: Crudely there is a compensation between
the propagator for the see-saw neutrinos and the factors coming from the 
Dirac mass matrix elements. This compensation is a phenomenon that can
very easily occur on general grounds, especially in a model version like
the most successful pattern which we use, the $\phi_{B-L}$ field has 
such quantum numbers that it is the matrix element number $(3,3)$ in the 
right-handed neutrino mass matrix which becomes the biggest one. 
Then namely as one goes away from the $(3,3)$ corner of this matrix 
the matrix elements of the right-handed mass matrix 
becomes smaller and smaller by factors corresponding to the 
successive need for more and more Higgs fields -- actually $\xi$ and $\chi$
-- to transfer the excess charges into those of the $(3,3)$-element 
in order to finally use the $\phi_{B-L}$. But now the Dirac matrix has 
it somewhat 
similarly also with $(3,3)$ matrix element dominating. It is even so that 
the extra charges to be gotten rid of when we compare matrix elements
deviating by having exchanged one right-handed neutrino (Weyl) field by the 
next is the same whether one makes this comparison in the right-handed 
neutrino matrix or in the Dirac one. The right-handed neutrinos of course
deviate in quantum numbers in the same way, whichever is the matrix we 
happen to consider.  There therefore can just come in a compensation 
between the propagator and the Dirac mass matrix factors. The 
compensations suggested here indeed take place in the case of 
the comparison of matrix elements with the $\nu_{R \mu}$ and 
the $\nu_{R e}$ (concepts that 
can be assigned meaning in our model because we have the proto-families).
This compensation then is one reason that we get the rather small
hierarchy ratio for the neutrinos as compared to the charged fermions.

The large mixing angle for atmospheric neutrino mixing is in our model 
achieved not so $100\%$ nicely by using the only for the neutrino
oscillations really relevant Higgs field $\chi$ as a parameter just 
to fit this big mixing angle, it comes by taking it of the same order
crudely as the already in the charged sector relevant Higgs field $T$.
Now because of this enforced choice it is then difficult to get 
a big mass ratio, a big hierarchy, for the two
in the atmospheric neutrino oscillation involved neutrino flavours.
The model actually only has factors that are of order unity in
principle to offer to this mass ratio, but it turns out that indeed the 
``factorial correction'' factors  and the way eigenvalues are computed is
sufficient to make the sufficiently big mass ratio.  One can say that in
principle the seemingly hierarchical mass square ratio is indeed 
not hierarchical, but only some factorial factors {\it etc}. 
in principle of order unity.

\section{Do we disfavour SUSY?}
\indent

In our model it was all the time the spirit that we did not include 
supersymmetry. However, at first one may ask if one could not very 
easily imagine the same model provided with SUSY-partners for all the 
particles in the model and postulate SUSY. At first it will also
be o.k. but there are some problems introduced by such a procedure: 
For one thing a realistic SUSY-model would presumably have gravity and
super gravity with gravitinos. The latter would get a mass being the 
geometrical mean of the SUSY-breaking scale and the Planck scale \ie\ about 
$3\times10^{9}~\GeV$. That however would now endanger the philosophy of our
model because we get the main contribution to the baryon asymmetry from
the decay of the heaviest of the three right-handed neutrinos and that 
neutrino has in our model 
a mass of the order of $10^{12}~\GeV$, and thus we cannot tolerate 
an inflation period as will be needed to avoid the gravitinos from surviving 
into the later times in the cosmological development of the universe. But
surviving gravitinos would not be allowed phenomenologically and our 
interpretation of the mechanism for the baryon asymmetry production 
could not work and our agreement would be an accident.

In this way supergravity is disfavoured from our model point 
of view.

There is another way in which SUSY is also disfavoured: %
\noindent
One of the predictions of our model is that order of magnitude-wise 
the up-quark and the electron (as well as with a correction of
the order $2$ or so the down-quark) have at the fundamental scale 
the same mass. In the SUSY models this prediction would necessarily be 
an accident because one in SUSY models one cannot use the same Higgs fields 
to provided the mass for the electron and for the up-quark. This is one
well-known reason why it is needed with two Weinberg-Salam Higgs fields in the 
SUSY models, another is to cancel anomalies of the Higgsinos.
But not only the Weinberg-Salam Higgs would have to be doubled in this way,
no, all the Higgs fields introduced into our model and relevant for
say the relation between the electron and up-masses, namely $W$, $T$, $\xi$ 
and $\phi_{WS}$ must be accompanied by another Higgs field to 
do their job of mass giving whenever the Hermitian conjugate field was 
used in the non-SUSY model. Thus taking our explanation for say the 
electron to up-quark mass ratio as a basically serious one it would be 
too accidental to have so many Higgs fields have the same expectation value as is needed in the SUSY model. Thus again SUSY is 
disfavoured because it would deprive us of major predictive power. 
Also in other cases than the electron to up-quark mass relation we would need
the doubling of the Higgses since for instance the mass of the charm-quark 
is produced by means of $W^{\dagger 2}$ while the dsb-quark masses use
$W$ (without dagger).

\section{Correlation of quantum numbers}
\indent

The success of our model with respect to fitting of the overall suppression 
scale crudely relative to the fluctuation from matrix element of the 
matrix elements in the mass matrices also confirmed by the success 
with baryon asymmetry could {\it a priori} be taken to mean that our model
should be correct w.r.t. its huge -- maximal -- number of approximately 
conserved quantum numbers, but really it is only the roughly 
speaking independently assigned quantum numbers that count in 
influencing the suppression checked. By the quantum numbers 
being independently assigned we here mean that one might be allowed 
to think of them as having been assigned in a random
way independently of each other to the various quark and 
lepton Weyl components.

It is, however, to be stressed that such an independence is 
far from true in our Anti-GUT model for a couple of reasons:
\begin{enumerate}
\item We have extended into our model the in fact rather remarkable 
feature of the system of quantum numbers in the Standard Model: 
Knowing the weak hyper charge $y/2$ you can always 
know the non-abelian representation (under $SU(2)$ and $SU(3)$) 
according to a rule that may be described by:
\begin{enumerate}
\item the charge quantisation rule: $d/2+t/3+y/2=0\;({\rm mod}\;1)\nn,$  
where $t$ is the triality of the $SU(3)$ representation, \ie\  
$t=1\; ({\rm mod} \;3) $ for quarks, $t=-1 \; ({\rm mod}\; 3)$ for 
anti-quarks, $t=0\; ({\rm mod}\; 3)$ for gluons, \etc\ and 
$d$ is the ``duality'' of the $SU(2)$representation meaning 
$d$ is even for integer weak isospin and odd for half integer 
weak isospin.
\item always take the smallest allowed representation, w.r.t. 
dimensionality say, for the non-abelian representations.
\end{enumerate}

Partly motivated by the observation \cite{Brene} that Nature 
seems to want so much the just mentioned charge quantisation 
rule to be as complicated as possible that it could be used 
as a principle telling us what the
Standard Model gauge group should be, we postulate 
in our model the rules $(a)$ and $(b)$ to work for each family
of gauge fields and representation assignments separately. Having 
done that we do not even need to put the 
non-abelian representations into the table \ref{table1} 
describing our model. It also means that 
once a chain of Higgs fields have been found providing 
the right exchange of family specific weak hypercharges
the corresponding family specific non-abelian quantum 
numbers are working with the same Higgs field VEV's.
Once we have postulated these rules $(a)$ and $(b)$ we have no 
further suppression from the non-abelian gauge symmetry,
it goes by itself just following the abelian quantum numbers as slaves.
\item Even the family specific $U(1)_{B-L, i}$ quantum 
numbers are restricted so much by the anomaly cancellation 
requirement that they also come to follow the weak 
hypercharges for the different families as slaves. At least 
we can see that they do indeed follow them systematically 
from the very definition $B-L$ for a family. You can namely easily 
give a rule to find the baryon number as well as the lepton number 
from the $y_i/2$ family specific weak hypercharge 
quantum numbers. 
\end{enumerate}

These relations between the various quantum numbers mean that 
there are roughly at most the three family specific weak hypercharges 
to make up the independent quantum numbers. So we can after 
this not expect the number of effectively independent 
charges to be more than $n=3$. Now actually our model has the 
field $S$ with almost vacuum expectation value unity 
in the fundamental units, and that means that actually the gauge 
group is in reality reduced to one family specific weak 
hypercharge less. This means that the number of independent 
charges in our model rather is $n=2$, just as the 
statistical estimates suggested as being the from the 
statistics of the charged fermion spectra preferred value.

The lesson from these considerations seems to be: %
\noindent
Even when one seeks with mild prejudices a model with as many as 
possible charges of the gauge type acting on the 
known quarks and leptons, especially the no anomaly constraints 
make them so correlated that the effective number 
of independently assigned charges is only about $3$, close to 
the number phenomenologically suggested. So one can hardly
get more than that, then correlations come up. 

Of course those quantum numbers which follow others as slaves 
are not really accessible by the mass matrix fitting to 
phenomenology and there is no evidence supporting their existence 
even if say our model fits, as it does, perfectly.

Only around two of the family specific weak hypercharges in our 
model could have any hope of having the support
from the successful phenomenology, the rest can only get their 
support to the extend that they were of some importance 
in via the anomaly constraints to restrict these family dependent 
weak hypercharge assignments.

\section{Conclusion}
We have presented a model in which it is assumed that:
\begin{itemize}
\item An under some condition maximal gauge group: $\AGUT$.
\item the set of fermion and scalar fields given by the table. 
\item All couplings and mass parameters are in fundamental (= Planck) 
units of order unity, so that only vacuum expectation values are 
of different order of magnitudes. The order of one couplings can 
be treated statistically (\ie\ as random numbers).
\end{itemize}

The number of measured quantities which are predictable with our 
model, quark and charged lepton masses and
its mixing angles containing Jarlskog triangle area $J_{CP}$ and also
two mass square differences for the neutrino and the three of 
their mixing angles, is $19$. Our model successfully predicted 
all these quantities using only six parameters\footnote{The VEV of 
Weinberg-Salam we have not counted as a parameter because of its
relation to the Fermi constant.}: the genuine number of predicted
parameters is thus $13$. But we have taken into the predictions 
the quantity $\tan^2\theta_{e3}$ for which CHOOZ has only upper bound.

We have suggested that the well fitting of this model with respect 
also to the question of the ratio of the spread in the spectrum 
of either matrix elements or simply masses of the charged quark 
and leptons relative to the average in a logarithmic way 
of the Yukawa couplings means that our model should have some 
truth in it w.r.t. the number of different  gauge charges 
participating effectively in suppressing the 
mass relative to the weak scale. 

Also the fitting of the baryon asymmetry should be a test of the 
well functioning of the model w.r.t. the general size -- logarithmic 
average say -- of the Yukawa couplings and so test also that the 
number of effectively used charge species is roughly o.k.

Since our model has -- if even counting the field $S$ which 
gives only tiny suppression as though having evidence in our 
fit for being there -- the maximal number of gauge charges, we 
may take our fit as supporting the thesis that in the true 
model there shall be many gauge charges! That is to say that 
the naive looking at the spectrum of quarks and leptons saying 
that it looks that we need rather many charges to be 
approximately conserved to explain the at first surprisingly low 
Yukawa couplings is correct.

However it must be admitted that we have 
actually starting from a model related in ideas to the 
present one constructed an example of a model with 
only two extra $U(1)$ charges \cite{Colin2} relative to the Standard Model. 
The quantum numbers in this model, however, are quite large and not 
like in the present model rather small, so the connection to 
fitting data from the number of charges is not but at best 
statistical, otherwise models with so different numbers 
of charges could not as in the example even match precisely 
in predictions. Really this model is -- in first approximation -- 
the present one with the decoration of the quantum numbers that follow 
the others as slaves thrown away.
 
Concerning the calculation of the baryon number asymmetry 
it turned out in our model that the main contribution to 
the baryon number came from the decay $CP$ violation of the 
heaviest one of the three see-saw particles because it had 
\begin{itemize}
\item only little dilution because its life time was very 
close in magnitude to the Hubble time scale at that time 
when this see-saw particle went out of equilibrium, while the 
two lighter see-saw particles decayed faster than their 
respective Hubble times. 
\item The middle mass see-saw particle and the heaviest had the 
same relative $CP$-asymmetry in their decays, while the lightest 
had an order of magnitude less.
\end{itemize}
If we for pedagogical reasons took it that the third 
(\ie\ the heaviest) right-handed neutrino contribution dominated, 
the picture was simply that the decay products -- after some 
wash out, which, however, due to the separate approximate 
conservation of three different family specific $(B-L)$-quantum 
numbers defined by association with the three right-handed 
neutrinos was not so tremendous -- from this heaviest of the 
see-saw neutrinos make up the excess of $(B-L)$.

\section*{Acknowledgements}

We are grateful to W.~Buchm{\"u}ller, C.D.~Froggatt, 
K.~Kainulainen, M.~Pl{\"u}macher, S.~Lola and T.~Yanagida
for useful discussions. We wish to thank 
the Theory Division of CERN for the hospitality extended to us
during our visits where part of this work was done.  H.B.N. 
thanks the EU commission for grants SCI-0430-C (TSTS), 
CHRX-CT-94-0621, INTAS-RFBR-95-0567 and INTAS 93-3316(ext). Y.T. 
thanks the Scandinavia-Japan Sasakawa foundation for grants No.00-22.



\begin{thebibliography}{99}
%
\bibitem{oldantiGUT1}
C.~D.~Froggatt, hep-ph/9908471.
%
\bibitem{oldantiGUT2}
C.~D.~Froggatt and H.~B.~Nielsen, hep-ph/9905445.
%
\bibitem{oldantiGUT3}
C.~D.~Froggatt, M.~Gibson, H.~B.~Nielsen and D.~J.~Smith, hep-ph/9810391.
%
\bibitem{oldantiGUT4}
C.~D.~Froggatt and H.~B.~Nielsen, Nucl.\ Phys.\ {\bf B147} (1979) 277.
%
\bibitem{oldantiGUT5}
C.~D.~Froggatt and H.~B.~Nielsen, hep-ph/9810388.
%
\bibitem{oldantiGUT6}
C.~D.~Froggatt and H.~B.~Nielsen, hep-ph/9710391.
%
\bibitem{oldantiGUT7}
D.~L.~Bennett, H.~B.~Nielsen and C.~D.~Froggatt, hep-ph/9707234.
%
\bibitem{oldantiGUT8}
C.~D.~Froggatt, H.~B.~Nielsen and D.~J.~Smith, hep-ph/9707205.
%
\bibitem{oldantiGUT9}
C.~D.~Froggatt, H.~B.~Nielsen and D.~J.~Smith, Phys.\ Lett.\ {\bf B385} (1996) 150
%
\bibitem{oldantiGUT10}
C.~D.~Froggatt, {\it Contribution to 27th International Conference on High Energy Physics (ICHEP), Glasgow, Scotland, 20-27 Jul 1994}.
%
\bibitem{oldantiGUT11}
C.~D.~Froggatt, G.~Lowe and H.~B.~Nielsen, Nucl.\ Phys.\ {\bf B420} (1994) 3.
%
\bibitem{oldantiGUT12}
C.~D.~Froggatt, G.~Lowe and H.~B.~Nielsen, Phys.\ Lett.\ {\bf B311} (1993) 163.
%
\bibitem{neutrinoAGUT1}
C.~D.~Froggatt, M.~Gibson and H.~B.~Nielsen, Phys.\ Lett.\ {\bf B409} (1997) 305
\bibitem{neutrinoAGUT2}
C.~D.~Froggatt, M.~Gibson, H.~B.~Nielsen and D.~J.~Smith, Int.\ J.\ Mod.\ Phys.\ {\bf A13} (1998) 5037.
%
\bibitem{NT1}
H.~B.~Nielsen and Y.~Takanishi, Nucl.\ Phys.\ {\bf B 588} (2000) 281.
%
\bibitem{NT2}
H.~B.~Nielsen and Y.~Takanishi, hep-ph/0011062.
%
\bibitem{NT3}
C.~D.~Froggatt, H.~B.~Nielsen and Y.~Takanishi, hep-ph/0011168.
%
\bibitem{NT4}
H.~B.~Nielsen and Y.~Takanishi, ``Baryogenesis via lepton number violation in Anti-GUT model,'' NBI-HE-00-48.
%
\bibitem{colinstatistical1}
C.~D.~Froggatt and H.~B.~Nielsen, Nucl.\ Phys.\ {\bf B164} (1980) 114.
%
\bibitem{colinstatistical2}
C.~D.~Froggatt and H.~B.~Nielsen, {\it Contribution to Int. Europhysics 
Conf. on High Energy Physics, Uppsala, Sweden, Jun 25 - Jul 1, 1987}.
%
\bibitem{colinstatistical3}
C.~D.~Froggatt and H.~B.~Nielsen, {\it  In Bergen 1979, Proceedings, 
Neutrino '79, Vol.1, 320-331}.
%
\bibitem{colinstatistical4}
C.~D.~Froggatt and H.~B.~Nielsen, Nucl.\ Phys.\ {\bf B147} (1979) 277.
%
\bibitem{DonPicek1}
D.~L.~Bennett and H.~B.~Nielsen, Int.\ J.\ Mod.\ Phys.\ {\bf A14} (1999) 3313.
%
\bibitem{DonPicek2}
D.~L.~Bennett and H.~B.~Nielsen,
``Constants of nature from a multiple point criticality fine tuning mechanism: Predictions of standard model gauge couplings,'' NBI-HE-94-45.
%
\bibitem{DonPicek3}
D.~L.~Bennett and H.~B.~Nielsen, Int.\ J.\ Mod.\ Phys.\ {\bf A9} (1994) 5155.
%
\bibitem{DonPicek4}
H.~B.~Nielsen, D.~L.~Bennett and I.~Picek, {\it Talk given at Workshop on Superstring Theory, Kanpur, India, Dec 13-24, 1987}.
%
\bibitem{DonPicek5}
H.~B.~Nielsen, D.~L.~Bennett and I.~Picek, {\it Talk given at Beijing Workshop on String Theories, Beijing, China, Jul 6 - Sep 5, 1987}.
%
\bibitem{DonPicek6}
D.~L.~Bennett, H.~B.~Nielsen and I.~Picek, Phys.\ Lett.\ {\bf B208} (1988) 275.
%
\bibitem{confusion1}
D.~L.~Bennett, N.~Brene, L.~Mizrachi and H.~B.~Nielsen, Phys.\ Lett.\ {\bf B178} (1986) 179.
%
\bibitem{confusion2}
D.~L.~Bennett, H.~B.~Nielsen, N.~Brene and L.~Mizrachi, 
{\it Talk given at 20th Int. Symp. on Theory of Elementary Particles, Ahrenshoop, Germany, Oct 13-17, 1986}.
%
\bibitem{confusion3}
H.~B.~Nielsen and D.~L.~Bennett,{\it Talk given at Conf. 
on Disordered Systems, Copenhagen, Denmark, Sep 1984}.
%
\bibitem{confusion4}
H.~B.~Nielsen and N.~Brene, Nucl.\ Phys.\ {\bf B224} (1983) 396.
%
\bibitem{Rajpoot}
S.~Rajpoot, Phys.\ Rev.\ D {\bf 24} (1981) 1890.
%
\bibitem{GS}
M.~B.~Green and J.~Schwarz, Phys.\ Lett.\ {\bf 149} (1984) 117.
%
\bibitem{Barshay}
S.~Barshay and G.~Kreyerhoff, ``Another calculation of the Higgs 
mass and the top mass from the principles of H.~B.~Nielsen: 
$m(H)\approx163~\GeV$ for  $M(t)\approx190~\GeV$'' PITHA-96-15.
%
\bibitem{FY}
M.~Fukugita and T.~Yanagida, Phys.\ Lett.\  {\bf B174} (1986) 45.
%
\bibitem{see-saw1}
T.~Yanagida, in Proceedings of the Workshop on Unified
Theories and Baryon Number in the Universe, Tsukuba, Japan (1979), eds.
O. Sawada and A. Sugamoto, KEK Report No. 79-18. 
\bibitem{see-saw2}
M.~Gell-Mann, P.~Ramond and R.~Slansky in Supergravity, 
Proceedings of the Workshop at
Stony Brook, NY (1979), eds. P.~van Nieuwenhuizen and D.~Freedman
(North-Holland, Amsterdam, 1979).
%
\bibitem{Luty}
M.~A.~Luty, Phys.\ Rev.\  {\bf D45} (1992) 455.
%
\bibitem{BuPlu1}
W.~Buchm{\"u}ller and M.~Pl{\"u}macher, Phys.\ Lett.\  {\bf B431} (1998) 354.

\bibitem{BuPlu2}
W.~Buchm{\"u}ller and M.~Pl{\"u}macher, hep-ph/0007176.
%
\bibitem{CRV}
L.~Covi, E.~Roulet and F.~Vissani, Phys.\ Lett.\  {\bf B384} (1996) 169.
%
\bibitem{KT}
E.~W.~Kolb and M.~S.~Turner, {\it The Early Universe}, %
Addison-Wesley, Redwood City, USA, 1990.
%
\bibitem{'tHooft1}
G.~'t Hooft, Phys.\ Rev.\ Lett.\  {\bf 37} (1976) 8.

\bibitem{'tHooft2}
G.~'t Hooft, Phys.\ Rev.\ {\bf D14} (1976) 3432. 
%
\bibitem{sphaleron}
V.~A.~Kuzmin, V.~A.~Rubakov and M.~E.~Shaposhnikov, 
Phys.\ Lett.\  {\bf B155} (1985) 36.
%
\bibitem{HT} 
J.~A.~Harvey and M.~S.~Turner, Phys.\ Rev.\ {\bf D42} (1990) 3344.
%
\bibitem{API}
A.~Pilaftsis, Int.\ J.\ Mod.\ Phys.\ {\bf A14} (1999) 1811.
%
\bibitem{Witten1}
M.~B.~Green, J.~H.~Schwarz and E.~Witten, ``Superstring Theory. 
Vol. 1: Introduction,'' {\it  Cambridge, Uk: Univ. Pr. (1987) 
469 P. (Cambridge Monographs On Mathematical Physics)}.

\bibitem{Witten2}
M.~B.~Green, J.~H.~Schwarz and E.~Witten, ``Superstring Theory. 
Vol. 2: Loop Amplitudes, Anomalies And Phenomenology,'' {\it  Cambridge, %
Uk: Univ. Pr. (1987) 596 P. (Cambridge Monographs On %
Mathematical Physics)}.

%
\bibitem{douglas}
C.D.~Froggatt, H.B.~Nielsen and D.J.~Smith, in progress.
%
\bibitem{totsuka}
Y.~Totsuka, talk at 8th International Conference on 
Supersymmetries in Physics (SUSY2K),  26 June - 1 July 2000, 
CERN, Geneva, Switzerland.
%
\bibitem{SK}
Y. Fukuda {\it et al.}, Super-Kamiokande Collaboration, 
Phys.\ Lett.\  {\bf B467} (1999) 185.
%
\bibitem{sobel}
H. Sobel, talk at  XIX International Conference on Neutrino Physics 
and Astrophysics, Sudbury, Canada, June 2000. 
%
\bibitem{toshito}
T. Toshito, 
talk at the XXXth International Conference on High Energy 
Physics, July 27 - August 2, 2000 (ICHEP 2000) Osaka, Japan.
%
\bibitem{suzuki} 
Y. Suzuki, talk  at XIX International Conference on Neutrino Physics 
and Astrophysics, Sudbury, Canada, June 2000.
%
\bibitem{takeuchi}
T. Takeuchi, talk 
at the XXXth International Conference on High Energy 
Physics, July 27 - August 2, 2000 (ICHEP 2000) Osaka, Japan.
%
\bibitem{chlorine1} B. T. Cleveland {\it et al.}, 
Astrophys. J. {\bf 496}, 505 (1998).
%
\bibitem{chlorine2}
R. Davis, Prog. Part. Nucl. Phys. {\bf 32}, 13 (1994).
%
\bibitem{chlorine3}
K. Lande, talk at XIX International Conference on Neutrino Physics and 
Astrophysics, Sudbury, Canada, June 2000. 
%
\bibitem{sage1} SAGE Collaboration, J. N. Abdurashitov {\it et al.},
Phys. Rev. {\bf C60}, 055801 (1999).
\bibitem{sage2}
V. Gavrin, talk at 
XIX International Conference on Neutrino Physics and Astrophysics,
Sudbury, Canada, June 2000.
%
\bibitem{gallex} GALLEX Collaboration, W.~Hampel {\it et al.},
Phys. Lett. {\bf B447} 127 (1999).
%
\bibitem{gno} E. Belloti, talk at XIX International 
Conference on Neutrino Physics and Astrophysics, Sudbury, 
Canada, June 2000.
%
\bibitem{BKS1}
J.~N.~Bahcall, P.~I.~Krastev and A.~Y.~Smirnov,
Phys.\ Lett.\  {\bf B477} (2000) 401.

\bibitem{BKS2}
J.~N.~Bahcall, P.~I.~Krastev and A.~Y.~Smirnov,
Phys.\ Rev.\  {\bf D62} (2000) 093004
%
\bibitem{valle}
M.C.~Gonzalez-Garcia, M.~Maltoni, C.~Pe\~{n}a-Garay 
and J.W.~Valle, hep-ph/0009350, to appear in Phys. Rev. {\bf D}.
%
%
%
%
\bibitem{Colin2}
C.~D.~Froggatt, M.~Gibson and H.~B.~Nielsen, Phys.\ Lett.\ {\bf B446} (1999) 256.
%
\bibitem{Brene}
H.~B.~Nielsen and N.~Brene, Nucl.\ Phys.\ {\bf B359} (1991) 406.



\end{thebibliography}
\end{document}